# Internet: Romania vs. Europe

**Graduated Student Papin Nicolae,
Assist.Prof. Tiberiu Marius Karnyanszky, PhD, Dipl.Eng., Dipl.Ec.,**
„Tibiscus" University of Timisoara, Romania

REZUMAT. În această lucrare sunt prezentate diferitele moduri de acces la Internet pentru utilizatorii casnici, atât pentru cei care sunt consumatori mici (timp consumat online sau valoare trafic lunar) mici sau mari (online nelimitat). Scopul principal al lucrării constă în realizarea unei comparații dintre situația Internetului în România fa'[ de alte țări din Europa cum ar fi Ungaria ( mai vestică decât România, deci puțin mai dezvoltată, dar totuși estică față de țările dezvoltate din vestul Europei, dar și țările cele mai dezvoltate ca Anglia, Italia, Franța, căt și cele în dezvoltare ca Polonia, căt și la periferia Europei cum este Ucraina.

## 1. Accessing the Internet

Whenever we open a newspaper, we turn on the TV or look into the postal box; we find a lot of promotions for the Internet providers (ISP – Internet Service Provider). This offers to the further clients the possibility to know and to select between a lot of competing companies and services.

The Internet is a gigantic network, widely crossed to connect individual elements, single computers or local networks, exchanging data using a single "language" called the TCP/IP protocol (Transmission Control Protocol / Internet Protocol).

Each individual computer is found on Internet using a single number, a single address that identifies that user. This number, called IP address, is extremely important for recognition of the computer into the network.

## 2. Users categories

**The minimal Internaut.** For everyone who is on-line less than 20 hours/month, the offers based on non-stop access (called Flat rate) or on a package with many hours are not interesting. In this case the offers based on the reduced volume of traffic (specified in MB) or on the reduced time (specified in hors) are reasonable. The transmission speed is reduced but it is not the most important parameter because the use of a fax-modem device cannot offer more speed.

The price, in the Western Europe, starts from 10 euro/month at 1 Mb/s, while in Romania the use of a fax-modem or of a CATV cable costs 9 $/month at the RDS operator and 12 euro/month at the UPC operator.

**The medium Internaut.** At 40-60 hours/month, the traffic speed is not the most important parameter but, in some cases, it disturbs the navigation. The non-stop access starting from 512 kb/s using a TV cable or DSL packages with included 4 GB traffic are offered at 20-40 euro/month in Western Europe. In Romania, prices are around 30 – 80 RON (8 – 22 euro) /month for a TV cable link or 100-150 RON (28 – 42 euro) /month for an ADSL at 512-1024 kb/s.

**The maximal Internaut.** In this case, the traffic is based on file downloads. The connecting time is over 100 hours/month and the traffic is more than 5 Gb. The most important parameter in this case is the connection speed, up to 16 Mb/s in Western Europe at around 50 euro/month.

**The content-based Internaut.** For the users that must be always online, a direct connection must be used. In Western Europe, it is offered by SDSL at more than 2,5 Mb/s downloads. Such alternative offers are based on TV cable for the business market at around hundreds of euros/month, even in Romania.

**The Internaut player.** For video games fans, the speed is not enough, the response time (ping) toward the on-line game servers is the most important. Especially for the 3D-Shooter league games this can lead to victory or defeat.

The best solution is the fiber connection (Fiber to the home) having only 30 ms latency around Europe. By satellite, WLAN, radio, it cannot be used, because these have around hundreds of ms latencies. The connection must be unlimited, at maximum 50-60 euros a month.

**The mobile type.** The exception is represented by the very mobile users, being all the time out of home or out of the office, using radio mobile services with Internet included, as the GSM with the GPRS technology at 128 kb/s, CSDM at 2,4 mb/s or UMTS and HSDPA, in the Europe, having 768 kb/s and 2,4 mb/s.

## 3. Romanian offers at the main mobile phone operators

**3.1.Connex / Vodafone** 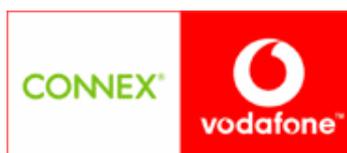 **www.vodafone.ro**

| Specification | Traffic included | Monthly fees | Access to | Extra traffic fees |
|---|---|---|---|---|
| Mobile Internet GPRS 10 | 10 MB | $3 | internet and wap | $0,3 / MB |
| Mobile Internet GPRS 50 | 50 MB | $10 | internet and wap | $0,3 / MB |
| Mobile Internet GPRS 100 | 100 MB | $15 | internet and wap | $0,3 / MB |
| Mobile Internet GPRS 250 | 250 MB | $25 | internet and wap | $0,3 / MB |

**The special Mobile Internet subscription** is created to offer connectivity to a laptop using a multifunctional card. It is activated by a special SIM card, dedicated only for the data transfer services but not for voice services.

**Wi-Fi.** Connex NetZone is the first Romanian commercial service that allows the transfer at broadband speed, using the Wi-Fi technology. This service is available inside the best hotels and continues its extinction.

**Broadband wireless access to the Internet.** Using a laptop and a simple Wireless LAN card, it is possible to navigate wireless all around the Connex NetZone is activated.*** The connection is very simple, because the access uses only a web interface, without any wires or plug. It permits sending and receiving mails with extra huge attachments or extra large files, especially when the transfer rate must be wide (videoconferences, databases etc).

**myX using a fax-modem.** After the Connex subscription is activated, every single user can access the Internet, tax-free. Connex Dial-up Free is based on the 10 hours/month at local taxes for the phone usage.

**Connex Dial-up Prepaid** offers the possibility to navigate for a few hours, the fees being paid before the usage ($4=16 hours, $5=20 hours, …, $25=117 hours). The speed is low – less than 56 kbps, this service being optimized for dial-up connections.

**Connex Dial-up Gold** offers two subscriptions:

| Specification | Monthly fees | Included hours | Supplementary access fees (/hour) | e-mail | web hosting | Connex subdomain |
|---|---|---|---|---|---|---|
| Dial-Up Hobby | 4 $ | 15 | 0,5 $ | 10 MB | 10 MB | Yes |
| Dial-Up Addict | 12 $ | Unlimited | - | 10 MB | 10 MB | Yes |

**Connex dial-up ISDN** is a data transfer service better than the dial-up usual connection. It permits very large data transfers, as in multimedia applications, and at the large broadband solution it can support, at the same time, a phone call and an Internet access. The monthly fees are $20 for 64 kbps or $32 for 128 kbps.

## 3.2. Orange  [www.orange.ro](www.orange.ro)

**Fax&Data Mobil** is a service package especially created to meet you communication necessities and you freedom of moving from a place to another. The Fax&Data package permits sending and receiving data files or fax messages of any type as well as to possess a Mobile E-mail, to navigate on internet and WAP at any time or any place.
   Benefits:
- Offers all the communication facilities necessary in an office or in the areas where there are no fix telephone lines.
- Eliminates the imposed barriers determined by the limited coverage of the fix telephone connection
- The SIM card for data transmissions or fax can be common or different form the SIM card for voice transmissions. Using a SIM card you have the possibility of speaking on the phone while you carry on data/fax transmissions or while you are connected to the Internet/Wap.
- Offers international access and roaming
   Charges: - Monthly subscription:
- In case of activating the same SIM with a voice number: 10USD/month.
- In case of activating a different SIM: 15USD/month
- The monthly subscription includes 50 notifications for the Mobile E-mail Standard Service

- For Professional Mobile E-mail 2 USD/month is added to the subscription fee.
- For the Professional Fax Message Box the subscription is the same with the Professional Vocal Message Box.
- Call charges:

| Specification | Internet (545) | | WAP (544) | Fax call/data Orange network | Fax call/data other networks |
|---|---|---|---|---|---|
| Price/minute (USD) | Peak | 0.06 | 0.09 | 0.10 | 0.15 |
| | Off peak | 0.01 | | | |

The transfer rate is 9600, 4800, 2400 bps depending on the receivers' capacities, while at receiving is of 9600 bps.

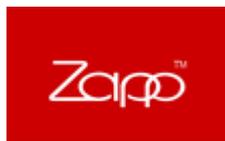

### 3.3. Zapp THE NEXT GENERATION NETWORK [www.zapp.ro](www.zapp.ro)

**Zapp Start**. Start is an option implicitly activated the very moment you have become a subscriber, charging as all the other data packages, for the time unit. If you constantly access the internet you can make the option for a dedicated data subscription including traffic minutes such as Upgrade Internet

**Zapp Turbo**. In addition Zap Turbo is free software for accelerating the speed of the Zap connection to Internet which improves up to ten times the data transfer rate (see Figure 1.). In other words, the average transfer speed is 200-300 kbps, sometimes reaching even 1500 kbps.

**HOTSPOT Zapp**. Read your e-mails and surf on Internet through a Zapp WiFi high speed broadband wireless connection. Hotspot Zapp combines the WiFi benefits with a speed up to 2.4 Mbps offered by Zapp Internet Express, a service based on CDMA2000 1xEV-DO technology.

All you need is to find yourself in a Zapp Hotspot-covered area having a laptop which:
- uses the Windows 2000 or Windows XP systems

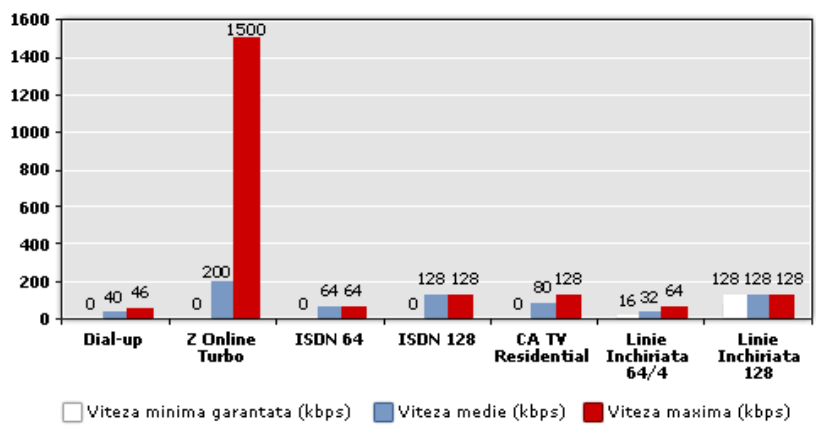

*Fig. 1:* Speed comparison

- be equipped with a WiFi Ethernet card or with a WiFi network PCMCIA/USB card concordant to IEEE 802.11b/g standard
- have Internet Explorer or any other browser installed

The access is free in order to test the service 10 minutes/day per computer (a laptop/PDA can get connected every day but not longer than 10 minutes) using the username "free" and the password "free".

In order to access this service on the following cards can be used:

| Offer | Zapp Internet Express 5 | Zapp Internet Express 15 | Zapp Internet Express 100 |
|---|---|---|---|
| Monthly subscription | 9 $ | 19 $ | 39 $ |
| Service package price Z010 | 59 $ * | 39 $ * | 39 $ * |
| Data traffic (EvDO or 1xRTT) | 5 hours/month ** | 15 hours/month ** | 100 hours/month ** |
| Additional traffic charges | 0.025 $ /minute | 0.02 $ /minute | 0.02 $ /minute |

**CDMA 1xEV-DO**. CDMA2000 1xEV-DO is built on an evolution of the CDMA2000 technology, optimized for the data transfer, using a carrier frequency independent from the voice one. Belonging to the wireless broadband class, CDMA2000 1xEV-DO offers a maximum transfer speed up to 2.4 Mbps for download and up to 153.4 kbps for upload (asymmetric transmission).

By introducing this new product on the market, Romania is the second country in Europe offering a commercial service based on CDMA1xEV-DO. The first service based on CDMA2000 1xEV-DO technology was launched in Chorea, in January 2002, using the frequency band of 800 MHz. In Europe, the first launching of this product took place in July, 2004, being offered by GSM Eurotel in Czech Republic, in the 450MHz frequency band. Nowadays, there are over CDMA1xEV-DO users worldwide. CDMA2000 1xEV-DO offers Internet surfing broadband speed, CDMA2000 1xEV-DO offers the most rapid mobile data transfer in Romania, with a maximum download speed of 2.4 Mbps.

Presently, worldwide there are 22 CDMA20001xEV-DO commercial networks in 16 countries, while in other 19 they are being implementing. Over 14 million subscribers use CDMA2000 1xEV-DO services, the specialists estimating that in 2008 this number will increase up to 120 million.

| **Specification** | **Zapp Internet Express Card 2** | **Zapp Internet Express Card 10** | **Zapp Internet Express CARD 30** | **Zapp Internet Express CARD 50** |
|---|---|---|---|---|
| Included credit | 2 $ | 10 $ | 30 $ | 50 $ |
| **Includes** | | | | |
| Data traffic(EvDO, 1xRTT or WiFi**) | 1 hour | 7 hours | 23 hours | 40 hours |
| ZappMobile Portal Access | ✓ | ✓ | ✓ | ✓ |
| **Validity** | | | | |
| Credit validity | 24 hours | 2 months | 4 months | 6 months |

## 4. The main similar offers in Europe

### 4.1. Germany

In Germany most of the offers are based on xDSL connections. The main suppliers have millions of customers:

159

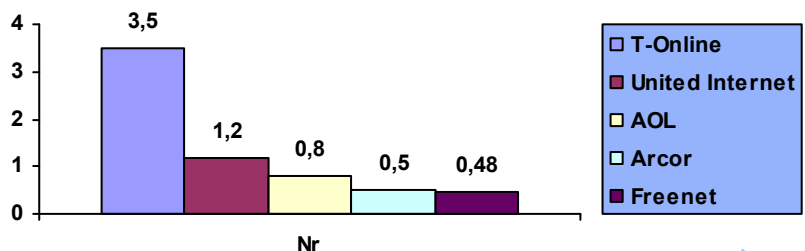

They have the following offer:

| Name | Offer | Down/up-stream min-max speed (kb/s) | Price Euro | Connection | Min.period months |
|------|-------|-------------------------------------|------------|------------|-------------------|
| 1&1 | 6 | 1024/128, 2048/192, 6016/576, 16000/1024 | 5, 10, 17, 20, 25, 30 | T-Com | 12 |
| 1click2surf | 6 volumic, 3 temporal, 3 unlimited | 1024-6016 | 1-50 | T-Com | 1 |
| AOL | 4 | 1024-16000 | 22-35 | T-Com | 12 |
| 3U Telecom | 2 | 6016 | 6-8 | T-Com | 12 |
| Arcor | 3 volumic | 16000/800 | 5+10 (1000) 15 (2000) 25 (6000) 35 (16000) | Arcor or T-Com | 12 |
| Congster | 2 | 6016/576 | 5,8,15 | T-Com | 12,12,3 |
| Freenet | 3 | 1024-6016 | 17,20, 25 | T-Com | 12 |
| Lycos | 2 | 6016 | 5,8 | T-Com | 12 |
| T-Online | 3.1 temporal 30 hours | 1024-6016 | 17+20+ 25 | T-Com | 12 |

For instance, T-Online from Germany has over 6 million active users (including those indirect connected by resellers), followed by the Italian ISP Tiscali (present in 15 countries in Europe) with 4.9 million and by the French ISP Wanadoo and AOL (9 countries in Europe) and 4 million users each.

## 4.2. Austria

The main providers are Aon, Inode, Eunet and UPC Telekabel.

**AON** is based on ADSL with the following offers:
- Volumic aonSpeed: with 400 MB – 20 Euro, with 800 MB – 30 Euro, with 2000 MB – 40 Euro, with 5000 MB – 50 Euro, at a speed of 1024/256 kbs, surpassing costing 7 cents/MB
- Unlimited traffic – 55 Euro/month, the maxim speed 2048/386 kbs, but limited at max 15 GB !!! After surpassing the access is stopped until the next month!!!; it is based on FUP Fair Use Prinzip (15 GB)
- Aon hotspot at 10 cents/minute.

**Inode** makes the following business offers:
- **Silber:** from 512/128 kbs to 5210/768 kbs between 29-109 Euro/month with a traffic limit of 1-60 GB, surpassing being charged with 4 cents/MB.
- **Gold:** from 512 to 18400 kbs with 99 Euro at unspecified figures: and as home-user offers:

| Name | Down/up-stream kbs speed | Traffic limit | Price: Euro/month |
|---|---|---|---|
| Small | 1024/256 | 2 | 30 |
| | 2048/512 | 5 | 39 |
| Medium | 3072/512 | 20 | 49 |
| | 4096/768 | 20 | 59 |
| Flat rate unlimited | +10 euro at small+medium | | |
| Large | 12288/1024 | 30 | 89 |

At Eunet, in collaboration with Ipass and the provided software the customer can get connected by his Internet account, even by WLAN hotspots with 11 Mbs (20.000 worldwide), from 40.000 locations on the Earth, meaning 160 countries.

The charges depend on the zones (areas on the earth) raging between 3 and 30 cents/minutes.

| Name | Down/up-stream kb/s speed | GB traffic limit | Price Euro/month |
|---|---|---|---|
| ADSL Flat | 5952/512 | unlimited | 315 |
| | 4096/512 | unlimited | 150 |
| ADSL Flex | 5952/512k | 30 GB | 125 |
| | 4096/512 | 10,30 | 90,114 |
| | 384/128 | 0.5 | 17 |

The ADSL connection can be done with or without a phone, depending on the choice; there are other offers based on Flat rate (no limit) or a traffic limit based on FUP or with an overuse charge of 1-7 cents/MB or with the disconnection until next month. Another variant is with volumic packages from 0.5 to 30 GB such as the above Flex.

**SDSL** presents an advantage for the small firms or natural persons as they can offer online services where a larger upload is necessary:

| Name | Down/up-stream kbs speed | GB taffic speed | Price Euro/month |
|---|---|---|---|
| Flex | 4096/4096 | 25 | 159 |
| | 2048/2048 | 20 | 138 |
| | 1024/1024 | 20 | 105 |
| | 768/768 | 15 | 85 |
| | 512/512 | 15 | 75 |
| | 256/256 | 15 | 55 |
| Flat | 4096/4096 | unlimited | 235 |
| | 2048/2048 | unlimited | 194 |
| | 1024/1024 | unlimited | 137 |
| | 768/768 | unlimited | 118 |
| | 512/512 | unlimited | 102 |
| | 256/256 | unlimited | 85 |

### 4.3. Hungaria

**Freestart** offer can be made in the fax-modem variant, with Freestart+ program at the charges: 1 month =790 Ft, 3 months =1790 Ft, 6 months =2290 Ft, and 12 months =5790 Ft.

| Name | Down/up-stream kb/s Speed | Minimum period – month contract | Price Ft/month |
|---|---|---|---|
| StartADSL | 512/96 | 18 | 6780 |
| StartADSL | 1024/128 | 18 | 9400 |
| StartADSL | 2048/192 | 18 | 11500 |
| StartADSL | 3008/384 | 18 | 13500 |

**T-com** offer is also based on using the fax-modem, which supposes charges according to the day period such as: 7-24 = 4.8 Ft/min, between 0-7 = 2.80 Ft/min.

| Name | Down/up-stream kb/s Speed | Minimum period – month contract | Price Ft/month |
|---|---|---|---|
| T-DSL Favorite | 1024/128 Kbit/s | 12,24 | 14780/13800 |
| T-DSL Favorite | 2048/192 Kbit/s | 12,24 | 16700/15700 |

For **Keystone** offer, there are the following conditions:

| Name | Down/up-stream kb/s Speed | Price Ft/month |
|---|---|---|
| ADSL Start | 512/96 | 4450 |
| ADSL Soho | 1024/128 | 5450 |
| ADSL Soho+ | 2048/192 | 6450 |
| ADSL Business II | 3008/384 | 16950 |
| ADSL Business III | 6144/512 | 29950 |

Similarly the **UPC** offer is:

| Name | Down/up-stream kbs Speed | Price Ft/month | Traffic limit GB/month |
|---|---|---|---|
| chello light | 512/128 | 5250 | 5 |
| chello classic | 2560/512 | 9500 | 30 |
| chello plus | 5150/512 | 12800 | 60 |

and after overusing, the traffic the download/ upload decreases to 64,256,512 kbs, depending on the subscription.

### 4.4. England

The main connections are through ADSL, but there are also offers through fax-modem used in the difficult accessible areas or by tourists, respectively connections through CATV (less, just in the cities), satellite (used more that CATV) and radio – WLAN.

The number of providers is immense; there are 10-25 providers in each connection type, each of them having at least 2 or 3 offers so that the offers number surpasses 300, in 4 main connection categories (fax-modem+ISDN, ADSL, CATV, satellite, WLAN).

The **euro1net** offer has the following options:

| Download kbs Speed | Price Pounds/month | Contract minimum period |
|---|---|---|
| 512 | 10 | 12,24 months |
| 1024 | 18,14,14 | 6,12,24 months |
| 2048 | 20,16,16 | 6,12,24 months |

while for **AOL** the options are:

| Download kb/s speed | Price pounds/month | Minimum contract period |
|---|---|---|
| Dial-up | 16 | - month |
| 1024 | 15 | 12 months |
| 2048 | 25 | 12 months |
| 2048 - radio | 30 | 12 months |

The best offer in England comes from **BE** with a down/up-load 24/1,3 Mbs speed with unlimited traffic, at 24 pounds. A comparison between **Be** and other providers is displayed below:

| Issue | Be unlimited | AOL Platinum | BT Option 4 | Wanadoo Heavy |
|---|---|---|---|---|
| Monthly price | £24 | £29.99 | £29.99 | £27.99 |
| Speed | 24 meg | 2 meg | 2 meg | 8 meg |
| Upload | 1.3 meg | 256 kbps | 256 kbps | 256 kbps |
| Limit | Unlimited | Unlimited | 40 gig | 30 gig |
| Faster than 512 kbps | 48x | 4x | 4x | 16x |
| Wireless modem | Free | Free | Free | £4 per month |

| Issue | Be lite | AOL Silver | BT Option 1 | Wanadoo Standard |
|---|---|---|---|---|
| Monthly price | £14 | £17.99 | £17.99 | £17.99 |
| Download | 24 meg | 1 meg | 2 meg | 8 meg |
| Faster than 1 meg | 24x | - | 2x | 8x |
| Traffic limit | 1 gig | Unlimited | 2 gig | 2 gig |
| Additional | £1 per gig | n/a | Upgrade only | Upgrade only |
| Wireless modem | Free | No | £25 supplement | £4 per month |

**Conclusions**

The offers for Internet access in **Europe** depend on the economic situation of a certain country, on the development level of the infrastructure, the number of the concurrent providers and their territorial coverage.

The biggest speeds are offered in countries like England, Italy, France, and Germany, with 24 Mbs, other countries being offered 2.4 Mbs or 5 Mbs.

Most of the access types are based on the old telephone infrastructure using ADSL, SDSL, ISDN or CATV. In the difficult accessible areas a fax-modem is used, WLAN, this being cheaper than the Satellite. If the Satellite is used the access can be cheaper by one-way.

The cheapest access way is by ADSL; in here the offers are many and attractive both as speed 0.5-24 Mbs and as price, 15-35 Euros. Thus, the best offers high speed Internet access is in **Italy, France and England, with 24 Mbs at 35 Euro, unlimited.**

In Romania the best offer is done by UPC with 1,5 Mbs – this being the maximum speed offered as well – yet at the price is quite high for the customer's income in this country, namely 26 Euro. UPC is planning to increase the speed up to 2.5 and 5 Mbs, starting with autumn, this way having the same offer as in Hungaria. Yet, there is great chance that by increasing the speed a limit of monthly traffic will be imposed, namely 30 and 60 GB/month, as in Hungaria.

The cheapest offer in Romania is done by RDS with 9 dollar for 0.5, while UPC charges 12 dollar for the same speed.